\documentclass[aps,twocolumn,prl,showpacs]{revtex4}
\usepackage{graphics}
\begin{document}
\title{Path finding strategies in scale-free networks}
\author {Beom Jun Kim}
\email{kim@tp.umu.se}
\affiliation {Department of Theoretical Physics,
Ume{\aa} University,
901 87 Ume{\aa}, Sweden}
\author {Chang No Yoon}
\author {Seung Kee Han} 
\affiliation{Department of Physics, Chungbuk National University, Cheongju, Chungbuk
361-763, Korea}
\author {Hawoong Jeong$^{\dagger}$}
\affiliation{Department of Physics, University of Notre Dame, Notre Dame,
Indiana 46556}
\preprint{\today}
\begin{abstract}
We numerically investigate the scale-free network model of
Barab{\'a}si and Albert 
[A. L. Barab{\'a}si and R. Albert, Science {\bf 286},  509  (1999)] 
through the use of various path finding 
strategies. In real networks, global network information is
not accessible to each vertex, and the actual path connecting two 
vertices can sometimes be much longer than the shortest one. 
A generalized diameter
depending on the actual path finding strategy is introduced, and 
a simple strategy,  which utilizes only local information on
the connectivity, is suggested and shown to yield small-world
behavior: the diameter $D$ of the network increases logarithmically with
the network size $N$, the same as is found with global strategy. 
If paths are sought at random, $D \sim N^{0.5}$ is found.
\end{abstract}


\maketitle

Complex network systems are abundant in many disciplines of sciences:
social networks composed of individuals interacting through
social connections, the world-wide web  where vertices are documents
and edges are hyperlinks to other documents, genetic networks
with genes or proteins connected by chemical interactions, and
economic networks, and so on~\cite{reviews}.
Topological information on such complex networks is becoming more and more
available through computer aided techniques, and
there is growing interest in the investigation of 
properties of such networks.

In complex networks, two remote vertices (or nodes) 
can be connected by many paths. 
Among those paths the length of the shortest one usually defines the ``distance'' $d_s(i,j)$
between the two vertices $(i,j)$, and the diameter $D_s$  of the network 
is defined to be either the maximum distance 
$D_s \equiv \max d_s(i,j)$, or the average distance 
$D_s \equiv \langle d_s(i,j)\rangle$.  We note that finding
the shortest path between two vertices requires global
information on how all vertices are interconnected, which
is usually not accessible to vertices that constitute the network. From this,
we suggest that the actual path finding strategy should be based only on
local information and should introduce the generalized 
distance $d(i,j)$ and diameter $D \equiv \langle d(i,j)\rangle$,
which depend on the actual path finding strategy. 
In Ref.~\cite{kleinberg}, a local path finding strategy
based on geometric information was suggested and 
the delivery time in Ref.~\cite{kleinberg} corresponds to
our generalized distance $d(i,j)$.

A recent study of the scale-free networks has revealed 
that the small-world phenomenon (that two distinct vertices can be
connected by a remarkably small number of intervening vertices)
emerges, i.e., $D_s \sim \log_{10}N$ with the network size $N$~\cite{diam}. 
However, only the shortest paths have been examined and the scaling
behavior can in principle be very different if a more realistic
path finding strategy is used.
In the experiment performed by Milgram~\cite{reviews,milgram} 
the person in Nebraska could not 
know what the shortest connection was to send a message
to the final recipient in Boston, but only tried to deliver it 
to his/her directly connected (on a first-name basis) friend who 
was supposed to be closer to the  person targeted in Boston.
Also, for the World-Wide Web (WWW), although two documents are just a few clicks
away if the shortest path is traced~\cite{diam}, 
it does not necessarily guarantee that one can get the information easily; 
sometimes it is possible to go around long distances to access a 
document which turns out to be only a few clicks away from the 
starting document.
Consequently, the small-world phenomenon observed in scale-free
networks through the use of the global path finding strategy 
still remains to be confirmed by use of a local strategy.
In this work we suggest several path finding strategies and show
that there indeed exists a very simple local strategy which results in
the small-world phenomenon.

We first construct the scale-free networks following the same method
used  in Ref.~\cite{BA} by Barab{\'a}si and Albert (BA): 
Starting with a small number ($m_0$) of vertices (or
nodes), a new vertex with $m$ edges (or links) is added at each time step 
in such a way that the probability $\Pi_i$ of being connected to the existing 
vertex $i$ is proportional to the connectivity $k_i$ 
(the number of vertices directly connected to $i$)
of that vertex, i.e., $\Pi_i = (k_i + 1)/ \sum_j (k_j + 1)$ with
summation over the whole network at a given instant~\cite{private}.
In Ref.~\cite{BA}, it has been shown that the above method
of constructing networks, which is composed of ideas of growth and preferential
attachment, results in the so-called scale-free networks, which shows 
the power-law behavior in the connectivity distribution.
Once the BA model network is constructed, we seek a path 
that connects two vertices in the network. At this stage, it is possible
to apply various strategies; for example, in previous studies~\cite{diam}, 
the shortest paths have been considered, the 
searching of which requires global 
information on the interconnections.
In real network systems like the WWW, the social network, and epidemic spread, it is very
unlikely that each vertex has global information and thus one
expects that each vertex has enough information only
for its directly connected vertices. In other words, the strategy used
when finding paths that connect two vertices should be based only on
local information not on global.

In this work we suggest three local strategies that one can use 
to find a path that connects two vertices:
(i) the vertex with the largest connectivity is tried first (maximum
connectivity first, MAX for short),
(ii) a vertex is chosen at random (random choice, RND), 
and (iii) the vertex with the larger connectivity has the higher probability 
to be chosen (preferential choice, PRF). We also
compare the results with the previously studied global strategy which seeks
the shortest path (SHT).
In the above experiment by Milgram, for example, the simplest and most intuitive
(and quite efficient indeed) strategy one can use is to first ask
the friend who has more friends than others, which corresponds
to the MAX strategy in this work. In reality, it is possible
that one does not know who has more friends than others, but even
in this case (s)he can try to first ask the friend who {\it seems} to have
more friends than others; this then has a close resemblance to
PRF strategy. The most naive strategy (it is not a strategy as
a matter of fact) will be asking anyone (s)he sees first, which corresponds
to RND strategy.

\begin{figure}
\centering{\resizebox*{!}{4.5cm}{\includegraphics{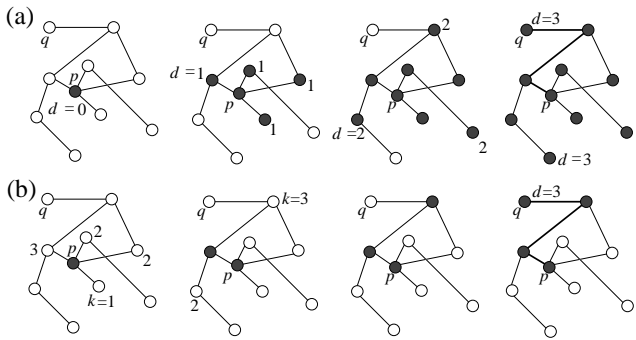}}}
\caption{Comparison of (a) SHT global strategy and
(b) local MAX strategy in finding a path connecting vertices
$p$ and $q$ (see the text for details). (a) For SHT,
the path-finding algorithm is similar to pouring water into
starting vertex $p$: water fills vertices with distance $d+1$
one step after it fills vertices with $d$. (b) For MAX, one proceeds
vertex by vertex by choosing the one with the largest connectivity
among directly connected vertices.}
\label{fig:strategy}
\end{figure}

Figures~\ref{fig:strategy}(a) and \ref{fig:strategy}(b), respectively,
show a comparison between SHT global strategy~\cite{foot:burning} and 
local MAX strategy when the path from vertex $p$ to
vertex $q$ is sought. In SHT, we give the distance as $d=0$ to vertex
$p$ [the first diagram in Fig.~\ref{fig:strategy}(a)],  
and $d=1$ is given to all vertices which are directly connected to $p$
[the second diagram in Fig.~\ref{fig:strategy}(a)]. Then the strategy proceeds to fill
all vertices with $d=2$ from vertices with $d=1$, and then $d=3$, and so on,
until $d$ values are given to all vertices in the network. According to this
SHT strategy the $d$ value attached to vertex $q$ is simply
its distance from $p$: In the example shown in Fig.~\ref{fig:strategy}(a),
one has $d=3$ for the path connecting $p$ and $q$, denoted by
the thick lines in the last diagram in Fig.~\ref{fig:strategy}(a). It is clear that
SHT determines the shortest path lengths but by using global
information.
Figure~\ref{fig:strategy}(b) explains local MAX strategy:
one starts at vertex $p$ and looks around its directly
connected vertices to find which vertex has the largest connectivity.
In Fig.~\ref{fig:strategy}(b), four vertices directly connected to $p$ have 
connectivities of
$k=1$, 2, 2, and 3, respectively, and thus we proceed to the vertex
with $k=3$ as shown in the second diagram. This strategy proceeds
until one arrives at the vertex that is directly connected to $q$.
In the example in Fig.~\ref{fig:strategy}(b) for  local MAX
strategy, one has the path denoted by the thick lines
in the last diagram of Fig.~\ref{fig:strategy}(b) with $d=3$.
The PRF strategy tries to proceed to the vertex with 
probability proportional to the connectivity, i.e., $P = k_i/\sum_j k_j$
with  summation over directly connected vertices of $i$. 
In the above example in Fig.~\ref{fig:strategy}(b), PRF connects 
to the vertex with $k=3$ [see the first diagram in Fig.~\ref{fig:strategy}(b)]
with probability $P = 3/(1+2+2+3) = 3/8$.
Of course, RND connects to a vertex at random at every stage in 
path finding.

\begin{figure}
\centering{\resizebox*{!}{6.0cm}{\includegraphics{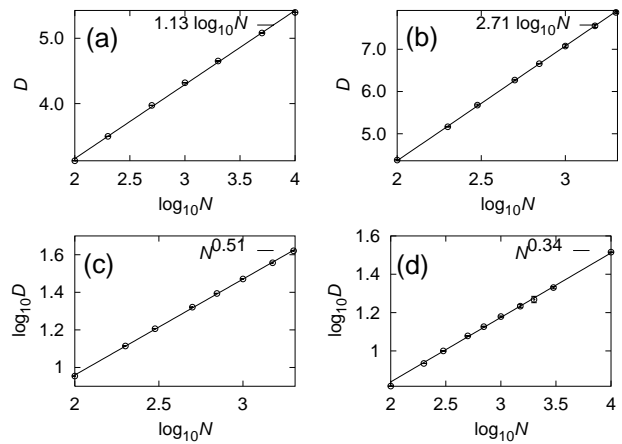}}}
\caption{
Diameter $D$ of the scale-free network vs size $N$.
(a) The shortest path (SHT) strategy, which requires global information
on interconnections, results in 
$D \sim \log_{10} N$~\protect\cite{diam}.
(b) Local MAX strategy also gives rise to the same
behavior as global SHT strategy, $D \sim \log_{10} N$,
although the $D$ is bigger than that of SHT. 
(c), (d) Local RND and PRF strategies 
are better described by the power-law behavior 
$D \sim N^{0.51}$ and $ \sim N^{0.34}$, respectively.
See the text for a description of each strategy.
} 
\label{fig:diam}
\end{figure}

The BA model network in this work is constructed with $m_0 = m = 2$
and the paths are found according to the above mentioned various
strategies. If the path contains self-crossing loops, we remove
them when distance $d$ is computed.
Figures~\ref{fig:diam}(a)--\ref{fig:diam}(b) display 
the scaling behavior of the generalized
diameter of the scale-free network for the strategies SHT, 
MAX, RND, and PRF, respectively. To get better statistics, the 
diameters were obtained from averages over many different
network realizations.
It is clearly shown that although the diameter determined
from local MAX strategy is a little bigger than that from
global SHT strategy and both exhibit the same small-world phenomenon, i.e., 
$D \sim \log_{10}N$ with network size $N$. 
On the other hand, RND strategy shows the power-law
behavior $D \sim N^{0.5}$~\cite{foot:MIN}.
Finally, PRF strategy shows the scaling behavior,
which fits well to  power-law $D \sim N^{0.34}$
[see Fig.~\ref{fig:diam}(d)].
As expected, the scaling behavior of the generalized diameter of the network 
depends not only on the structure of the
networks but also on the strategy used in path finding.
It should also be noted that a very simple MAX strategy, based on
local information instead of global, can result in the
same small-world effect as global SHT strategy.
On the other hand, we are puzzled by the fact that PRF and MAX strategies
show very different scaling behaviors although both
look quite similar. To compare both strategies in a more careful way,
one can change the probability used by PRF, e.g., 
$P \propto k^\alpha$ with $\alpha > 1$, 
to suppress the probability of the vertex with the lower connectivity 
being chosen.

For comparison, we also study the scaling behavior of the diameter
of the small-world network model by Watts and Strogatz (WS)~\cite{reviews}.
In the WS model, each vertex is almost equal to each other and there
does not exist a governing dominant vertex with very large connectivity.
This feature makes our local MAX and PRF strategies inefficient and
from numerical investigations we find that none of the local strategies tried
in this work yields the small-world phenomenon but gives
$D \sim N^a$ with $a \approx 0.6$ unanimously (we have used
the range of local connections $K=3$ and the rewiring probability $P=0.1$;
see Watts Strogatz's work in Ref.~\cite{reviews} for details of the model).
This is in a sharp contrast to the result from the SHT, which reveals
$D \sim \log_{10} N$.  From the above observation, we conclude
that the success of strategy MAX for the BA model
is due to the existence of the highly connected vertices, thus reflecting
the scale-free nature of the network.
It is of note that in Ref.~\cite{kleinberg} an efficient
local path finding strategy based on geometric informations was 
suggested for a two-dimensional small-world network model: 
When the path from vertex A to target B
is found, the strategy first connects C which is closest to B 
in a geometric sense among the A's directly connected vertices. 
We believe that our local strategies in this work
have some advantage when more abstract networks like the 
Internet and WWW are involved: 
In such networks vertices do not have coordinates and 
path finding strategies cannot use geometric information.

\begin{figure}
\centering{\resizebox*{!}{6.0cm}{\includegraphics{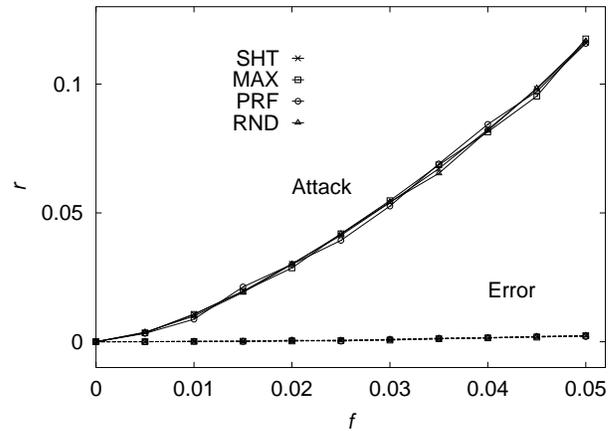}}}
\caption{
Ratio $r$ of failed connections vs the vertex failure fraction 
$f$[ = (number of vertices removed by errors or attacks)/(number of vertices)].
The error tolerance and the attack vulnerability of the scale-free
networks do not depend on the path-finding strategy. The network size,
$N = 1000$, and more than 100 different network realizations were
used to compute average values for each strategy.}
\label{fig:rtol}
\end{figure}
\begin{figure}
\centering{\resizebox*{!}{6.0cm}{\includegraphics{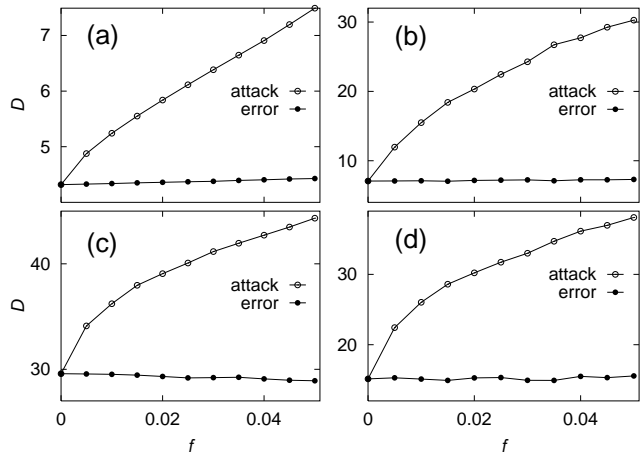}}}
\caption{
Error tolerance and attack vulnerability reflected in the diameter
of the network for strategies (a) SHT, (b) MAX, (c) RND, and (d) PRF
(see the text for details). All strategies show similar behavior as
implied by Fig.~\ref{fig:rtol} except that $C$ decreases
rather than increases for RND in the case of failures due to errors.
$N=1000$ and more than 100 different network realizations were 
used.}
\label{fig:dtol}
\end{figure}

We finally examine the error and attack tolerance of the BA model
(see Ref.~\cite{albert:attack} for a study based on  
SHT strategy) that is subjected to the various strategies.
We first remove $N'$ vertices in the network either randomly 
(error) or starting from the vertex with the highest connectivity 
(attack)~\cite{albert:attack},
and then measure how many pairs of vertices are connected by a
given path finding strategy. More specifically, we pick
two vertices, $p$ and $q$, and examine if $q$ can be connected
to $p$ by the path finding strategy given. Since there are
a total of $N(N-1)$ possible choices for the pair $(p,q)$ in
a network of size $N$,
the ratio $r$ of failed connections in Fig.~\ref{fig:rtol} is then defined as
\begin{equation} \label{eq:r}
r \equiv \frac{ \mbox{(number of failed connections)} }{ N(N-1) }.
\end{equation}
The tolerance in this article is detected either by $r$ in Eq.~(\ref{eq:r})
or by the change in diameter $d$ of the network. 
The former should not depend on the strategy since the 
strategy chosen can only change the path lengths between vertices once 
they belong to the same cluster.  
As expected, $r$ versus the failure fraction $f \equiv N'/N$ 
(the ratio between the number of vertices removed and the total 
number of vertices in the network) in Fig.~\ref{fig:rtol} shows the behavior
independent of the strategy. This robustness to the path finding
strategy implies that the error tolerance and the attack
vulnerability found in Ref.~\cite{albert:attack} are genuine 
topological characteristics of a scale-free network.
Figures~\ref{fig:dtol}(a)--\ref{fig:dtol}(d) confirm the similar tolerance behavior
reflected in diameter $d$ to errors and attacks for various strategies:
SHT, MAX, RND, and PRF, respectively.
For RND strategy in Fig.~\ref{fig:dtol}(c), it can be seen 
that removal of randomly
chosen vertices makes $d$ smaller as $f$ is increased from zero; this somewhat 
interesting behavior
is due to the fact that in RND path finding can be more 
efficient if unimportant connections are removed.

In summary, we have investigated the possibility of 
the small-world phenomenon in the scale-free network subjected to 
several local strategies of path finding. It was found
that there is  a very simple local strategy based on 
local connectivity information which leads to the scaling behavior
$D \sim \log_{10} N$. 
The error tolerance and the attack
vulnerability found previously by global strategy 
have been shown to be generic topological properties of
scale-free networks, and do not depend on the path finding
strategy.

\begin{acknowledgments}
S.K.H. and B.J.K. acknowledge support from the BK 21 project of
the Ministry of Education in Korea.
This work was supported in part by the Swedish Natural Research Council
through Contract No. F5102-659/2001 (B.J.K.), and by the
NSF through Grant No. PHY-9988674 and CAREER DMR97-01998 (H.J.). 
\end{acknowledgments}

{\it Note Added in Proof}: After submission of this paper we found
that the similar local search algorithm in networks with power-law
connectivity distribution was studied in Ref.~\cite{huberman}.


\begin{thebibliography}{10}

\bibitem[$\dagger$]{} Present address: Department of Physics, Korea Advanced Institute of Science and Technology, Taejon 305-701, Korea

\bibitem{reviews}
For reviews, see, e.g.,  
a special issue on complex systems in Science {\bf 284}, 79 (1999);
M.~E.~J. Newman, J. Stat. Phys. {\bf 101},  819  (2000);
D.~J. Watts, {\em Small Worlds} (Princeton University Press, Princeton, NJ, 1999);
S.~H. Strogatz, Nature (London) {\bf 410},  268  (2001), and references in
Refs.~\cite{diam} and \cite{BA}.

\bibitem{kleinberg} J.~M. Kleinberg, Nature (London) {\bf 406}, 845 (2000).

\bibitem{diam}
R. Albert, H. Jeong, and A.-L. Barab{\'a}si, Nature (London) {\bf 401},  130  (1999);
A.-L. Barab{\'a}si, R. Albert, and H. Jeong, Physica A {\bf 281},  69  (2000).



\bibitem{milgram}
S. Milgram, Psychology Today {\bf 2},  60  (1967).

\bibitem{BA}
A.-L. Barab{\'a}si and R. Albert, Science {\bf 286},  509  (1999);
A.-L. Barab{\'a}si, R. Albert, and H. Jeong, Physica A {\bf 272},  173  (1999).

\bibitem{private}
The use of $k_i + 1$ instead of $k_i$ makes it possible for vertices with zero
  connectivity to be linked to the network as time progresses.

\bibitem{foot:burning} This algorithm is called ``the burning algorithm''
or ``the breadth-first search algorithm;'' see, e.g., H.~J. Herrmann, D.~C. Hong, and H.~E. Stanley, J. Phys. A {\bf 17}, L261 (1984); M.~E.~J. Newman, Phys. Rev. E {\bf 64} 016132 (2001), and references therein.

\bibitem{foot:MIN}
We have also tested the MIN strategy, which is opposite to MAX and tries
  to first connect the vertex with the smallest connectivity. Through the use of
  MIN, which is the worst one can imagine among local strategies based on 
  connectivity, diameter $d$ is found to scale with the network size as $D
  \sim N^{0.53}$.

\bibitem{albert:attack}
R. Albert, H. Jeong, and A.-L. Barab{\'a}si, Nature (London) {\bf 406},  378  (2000).

\bibitem{huberman}
L.~A. Adamic, R.~M. Lukose, A.~R. Puniyani, and B.~A. Huberman,
   Phys. Rev. E {\bf 64} 046135 (2001).

\end{thebibliography}

\end{document}